%% file: paper.tex
\RequirePackage{lineno}

\documentclass[12pt]{iopart}

\usepackage{subfigure}
\usepackage{mathrsfs}
\usepackage{graphicx}
\usepackage{rotating}
\usepackage{xspace}
\usepackage{slashed}
\usepackage{multirow}
\usepackage{array}
\usepackage{amssymb}
\usepackage{cite}
\usepackage[hyperindex,breaklinks]{hyperref} 
\usepackage{atlasphysics}
\usepackage{preprintcover}  

\makeatletter 
\renewcommand{\p@subfigure}{} 
\renewcommand{\@thesubfigure}{\thesubfigure\hskip\subfiglabelskip} 
\makeatother 

\PreprintCoverPaperTitle{Search for excited electrons and muons in $\boldmath{\sqrt{s}=8}$~TeV proton--proton collisions with the ATLAS detector} 

\PreprintIdNumber{CERN-PH-EP-2013-131}  

\PreprintCoverAbstract{The ATLAS detector at the LHC is used to search for excited electrons and excited muons 
in the channel $pp\to \ell\ell^* \to \ell\ell\gamma$, assuming that excited leptons are produced via contact interactions. 
The analysis is based on 13~\ifb\ of $pp$ collisions at a centre-of-mass energy of 8~\tev. No evidence for excited leptons is found, and 
a limit is set at the 95\% credibility level on the cross section times branching ratio as a function of the excited-lepton mass $m_{\lstar}$.
For $m_{\ell^*}\geq 0.8$~\tev, the respective upper limits on $\sigma B(\lstar \rightarrow \ell\gamma)$ are 0.75~fb and 0.90~fb for the $e^*$ and $\mu^*$ searches.
Limits on $\sigma B$ are converted into lower bounds on the compositeness scale $\Lambda$.
In the special case where $\Lambda = m_{\lstar}$, excited-electron and
excited-muon masses below 2.2~\tev\ are excluded. }  

\PreprintJournalName{New J. Phys.}  

\def\lstar{\ensuremath{{\ell^{*}}}}
\def\zpjet{\ensuremath{Z +{\rm jets}}}
\def\zpgam{\ensuremath{Z + \gamma}}
\def\wpjet{\ensuremath{W+{\rm jets}}}
\def\ptmu{\ensuremath{\pt^{\mu}}}
\def\pte{\ensuremath{\pt^{e}}} 
 
\def\ptgam{\ensuremath{\pt^{\gamma}}}

\def\eff{\ensuremath{\varepsilon}}

\def\ifb{\mbox{fb$^{-1}$}}
\def\mcatnlo{MC@NLO}
\def\pythia{{\sc Pythia}}
\def\powheg{{\sc Powheg}}
\def\sherpa{{\sc Sherpa}}
\def\herwig{{\sc Herwig}}
\def\jimmy{{\sc Jimmy}}
\def\alpgen{{\sc Alpgen}}
\def\comphep{{\sc CompHEP}}

\def\geant{{\sc Geant}}
\def\photos{{\sc Photos}}
\def\medium{{\it medium}}
\def\tight{{\it tight}}

\begin{document}

\title[]{Search for excited electrons and muons in $\sqrt{s}=8$~TeV proton--proton collisions with the ATLAS detector}

\author{The ATLAS Collaboration}

\begin{abstract}
The ATLAS detector at the LHC is used to search for excited electrons and excited muons 
in the channel $pp\to \ell\ell^* \to \ell\ell\gamma$, assuming that excited leptons are produced via contact interactions. 
The analysis is based on 13~\ifb\ of $pp$ collisions at a centre-of-mass energy of 8~\tev. No evidence for excited leptons is found, and 
a limit is set at the 95\% credibility level on the cross section times branching ratio as a function of the excited-lepton mass $m_{\lstar}$.
For $m_{\ell^*}\geq 0.8$~\tev, the respective upper limits on $\sigma B(\lstar \rightarrow \ell\gamma)$ are 0.75~fb and 0.90~fb for the $e^*$ and $\mu^*$ searches.
Limits on $\sigma B$ are converted into lower bounds on the compositeness scale $\Lambda$.
In the special case where $\Lambda = m_{\lstar}$, excited-electron and
excited-muon masses below 2.2~\tev\ are excluded. 

\end{abstract}

\hyphenation{ATLAS}


\section{Introduction}


Although the Standard Model (SM) of particle physics is very successful at describing a large range of phenomena, it does not provide an explanation 
for the generational structure and mass hierarchy of quarks and leptons. 		 	 
Fermion compositeness models~\cite{patisalam1,patisalam2,eichten,cabbibo,hagiwara,BaurSpiraZerwas} aim at reducing the number of fundamental matter constituents
by describing SM fermions as bound states of more-elementary particles.
The existence of excited states would then be a direct consequence of the fermion substructure.

This paper reports on searches for excited electrons ($e^*$) and excited muons ($\mu^*$) using 13~\ifb\ of $pp$ collision data at a centre-of-mass energy of $\sqrt{s} =8$~\tev\
recorded in 2012 with the ATLAS detector at the Large Hadron Collider (LHC). 
Searches are based on a benchmark model~\cite{BaurSpiraZerwas} that describes excited-fermion interactions using an effective Lagrangian.
Excited leptons ($\lstar$) would be predominantly produced via four-fermion contact interactions, 
and are expected to decay into a lepton and a gauge boson, or a lepton and a pair of fermions. 
All unknown couplings of the model are set as in~\cite{BaurSpiraZerwas}. 
The contact interaction is then described by the Lagrangian
\[
 {\cal L}_{\rm contact} =\frac{2\pi}{\Lambda^2} j^\mu j_\mu\quad,\quad j_\mu = \overline{f}_L\gamma_\mu f_L +  \overline{f}_L^\ast\gamma_\mu f_L^\ast + \overline{f}_L^\ast\gamma_\mu f_L + h.c. 
\]
where $\Lambda$ is the compositeness scale, $j_\mu$ is the fermion current for ground states ($f$) and excited states ($f^*$), and ``$h.c.$" stands for Hermitian conjugate. 
The gauge-mediated decays are given by the Lagrangian
\[
{\cal L}_{\rm GM} = \frac{1}{2\Lambda} \overline{\ell}_R^\ast \sigma^{\mu\nu}\left[g 
\frac{\tau^a}{2}W_{\mu\nu}^a+g'\frac{Y}{2}B_{\mu\nu} \right] \ell_L +  h.c.
\]
where $\ell$ and $\ell^*$ are the lepton and excited-lepton fields, respectively, $W_{\mu\nu}$ and $B_{\mu\nu}$ are the $SU(2)_L$ and $U(1)_Y$ field-strength tensors,
and $g$ and $g'$ are the corresponding gauge couplings.
The searches described here focus on the single-production mechanism ($q\overline{q}\rightarrow {\lstar}^\pm{\ell}^\mp$) 
and the electromagnetic radiative decay mode $\ell^*\to\ell\gamma$.
The signature thus consists of events containing two same-flavour, opposite-charge leptons and a photon ($\ell^+\ell^-\gamma$ final state).  
The kinematic properties of the signal are determined by 
the excited-lepton mass ($m_{\ell^*}$) and the compositeness scale ($\Lambda$).
Due to unitarity constraints on contact interactions~\cite{BaurSpiraZerwas,unitarity}, the model does not apply in the regime $m_{\ell^*}>\Lambda$.
For most of the parameter space, the presence of excited leptons would appear as a peak in the lepton--photon mass spectrum. 
However, for $m_{\ell^*}\simeq\Lambda$, the width of the resonance can become significantly larger than the experimental mass resolution of the detector.
To avoid this complication as well as the lepton--photon pairing ambiguity, a search is performed for an excess in the $\ell \ell\gamma$ invariant mass ($m_{\ell\ell\gamma}$) spectrum.

Previous searches at LEP~\cite{LEPresults1,LEPresults2,LEPresults3,LEPresults4}, HERA~\cite{HERAresults1,HERAresults2} and the 
Tevatron~\cite{TEVATRONresults1,TEVATRONresults2,TEVATRONresults3,TEVATRONresults4} 
have found no evidence for excited leptons.  For the case where $m_{\ell^*}=\Lambda$, $e^*$ and $\mu^*$ masses below 
1.9~\tev\ have been excluded by both the ATLAS~\cite{lstar2011} and CMS~\cite{lstarCMS} experiments using $\sqrt{s}=7$~\tev\ data.


\section{ATLAS detector}


The ATLAS detector~\cite{atlas:detector} consists of an inner tracking system surrounded by a superconducting solenoid,
electromagnetic and hadronic calorimeters, and a muon spectrometer. 
It has a forward-backward symmetric cylindrical geometry\footnote[1]{ATLAS uses a right-handed coordinate system with the $z$-axis along the beam pipe. The $x$-axis points to the 
centre of the LHC ring, and the $y$-axis points upward. Cylindrical coordinates $(r,\phi)$ are used in the
transverse plane, $\phi$ being the azimuthal angle. The pseudorapidity is defined in terms of the polar angle $\theta$ as 
$\eta=-\ln\tan(\theta/2)$.} and nearly 4$\pi$ coverage in solid angle. 
Charged-particle tracks and vertices are reconstructed in silicon-based pixel and microstrip tracking 
detectors that cover $|\eta| <$ 2.5 and transition radiation detectors extending to $|\eta| <$ 2.0.
A hermetic calorimeter system, which covers $|\eta|$ $<$ 4.9, surrounds the 
solenoid. The liquid-argon electromagnetic calorimeter, which plays an important role 
in electron and photon identification and measurement, is finely segmented  
for $|\eta| < 2.5$ to provide excellent energy and position resolution. Hadron calorimetry is 
provided by an iron--scintillator tile calorimeter in the central pseudorapidity range $|\eta| < 1.7$ and a 
liquid-argon calorimeter with copper or tungsten as absorber material in the pseudorapidity range $1.5 < |\eta| < 4.9$. 
A spectrometer is installed outside the calorimeter to identify muons and measure their momenta with high precision. 
The toroidal magnetic field of the muon spectrometer is provided by three air-core superconducting magnet systems: one for the barrel and one per endcap, each composed of 
eight coils.  Three layers of drift-tube chambers and/or cathode-strip chambers provide precise coordinate measurement in the bending plane ($r$--$z$)
 in the region $|\eta |< 2.7$.  A system consisting of resistive-plate 
chambers for $|\eta| < 1.05$ and thin-gap chambers for $1.05<|\eta| <2.7$ provides measurement of the $\phi$ coordinate.
It also provides triggering capability up to $|\eta| = 2.4$.


\section{Simulated samples}


The simulation of the excited-lepton signal is based on calculations from~\cite{BaurSpiraZerwas}. 
Signal samples are generated at leading order (LO) with \comphep\ 4.5.1~\cite{comphep} using
MSTW2008 LO~\cite{mstw} parton distribution functions (PDFs). \comphep\  is interfaced with 
\pythia\ version 8~\cite{ComphepPythia,pythia} for the simulation of parton showers and hadronization.
The emission of photons via initial-state radiation and final-state radiation (FSR) is handled by \pythia.  
Only the single production of excited leptons followed by a $\ell^*\to\ell\gamma$ decay is simulated. 

For both the $e^*$ and $\mu^*$ searches, the dominant background arises from Drell--Yan 
processes accompanied by either a prompt photon from initial- or final-state radiation (\zpgam) or 
a jet misidentified as a photon (\zpjet).  The \zpgam\ background results in the same 
final state as the signal, whereas the \zpjet\ background is suppressed by imposing stringent 
requirements on the quality and isolation of the reconstructed photon. 
Small contributions from \ttbar\ and diboson ($WW$, $WZ$ and $ZZ$) production are also present in both channels. 
In the electron channel, the $W+\gamma+\rm jets$ background contributes to the $ee\gamma$ selection when a jet is misidentified as an electron.
Backgrounds from \wpjet\ and multi-jet events, including semileptonic decays of heavy-flavour hadrons, are suppressed by 
requiring the leptons and the photon to be isolated, 
and have negligible contribution to the total background after selection. 

The \zpgam\ sample is generated with \sherpa\ 1.4.1~\cite{SherpaManual} using CT10~\cite{ct10} PDFs
and includes the LO emission of up to three partons in the initial state. 
To avoid phase-space regions where matrix elements 
diverge, the angular separation between the photon and each lepton is required to be $\Delta R(\ell,\gamma) = 
\sqrt{(\Delta \eta)^2 + (\Delta \phi)^2}> 0.1$ and the transverse momentum of the photon (\ptgam) is 
required to be above 10~\gev.  
To ensure adequate statistics, 1.2 million events (equivalent to 37~\ifb) were generated for each of the electron and muon channels. 

The \zpjet\ and $W+\gamma+\rm jets$ backgrounds are generated with 
\alpgen\ 2.13~\cite{Mangano:2002ea} using CTEQ6L1~\cite{cteq6l1} PDFs. The 
\ttbar\ background is produced with \mcatnlo\ 3.41~\cite{mcatnlo} with CT10 PDFs.
In both cases, \jimmy\ 4.31~\cite{jimmy} is used to describe multiple parton interactions and \herwig\ 6.510~\cite{herwig} 
is used to simulate the remaining underlying event, parton showers and hadronization.  
The diboson processes are generated with \powheg~\cite{powhegdib} and \pythia\ using CT10 PDFs. 
For all these samples, FSR is handled by \photos~\cite{fsr_ref}.  
To remove overlaps between the \zpjet\ and \zpgam\ samples, 
\zpjet\ events with prompt photons are rejected if $\ptgam > 10$~\gev\ and $\Delta R(\ell,\gamma) > 0.1$.  
The predictions for \zpjet\ and $W+\gamma+\rm jets$ backgrounds are normalized using the data-driven techniques described in section \ref{sec:bkg}.  
Cross sections for diboson processes are evaluated at next-to-leading order~\cite{diboson} and 
the \ttbar\ cross section is calculated at approximate-next-to-next-to-leading order~\cite{ttbar}, 
with uncertainties of 5\% and $^{+10\%}_{-9\%}$, respectively.

The generated samples are processed using a detailed detector simulation~\cite{atlas:sim} based on 
\geant4~\cite{geant} to propagate the particles and account for the detector response.  
Monte Carlo (MC) minimum-bias events are overlaid on both the signal and background processes to simulate 
the effect of additional $pp$ collisions (pile-up).  Simulated events are weighted so that the distribution of the expected 
number of interactions per event agrees with the data, with an average of 20 interactions per bunch crossing.


\section{Data and selection}
\label{sec:selection}


The data were collected between April and October 2012 during stable-beam periods of $\sqrt{s} =8$~\tev\ $pp$ collisions, and 
correspond to an integrated luminosity of 13.0~\ifb\ for the electron channel and 12.8~\ifb\ for the muon channel~\cite{lumi7TeV}.
For the $e^*$ search, a trigger relying only on calorimetric information is used to select events.
It requires two electromagnetic clusters 
with transverse momentum (\pt) thresholds of 35~\gev\ and 25~\gev\ for the leading and subleading clusters, respectively,
with loose shower-shape requirements aiming to select electrons and photons. 
For the $\mu^*$ search, a single-muon trigger is used.
It requires a track to be reconstructed in both the muon spectrometer and the inner detector
with a combined track $\pt > 24$~\gev. 

Offline, events are selected if they contain at least two lepton candidates and a photon candidate.
A primary vertex with at least three associated charged-particle 
tracks with $\pt > 0.4$~\gev\ is also required. If several vertices fulfill this requirement, 
the vertex with the largest $\Sigma \pt^2$ is selected, where the sum is over all reconstructed tracks associated with the vertex.

Each electron candidate is formed from a cluster of cells in 
the electromagnetic calorimeter associated with a charged-particle track in the inner detector. 
For the $e^*$ search, two electron candidates are required. Their transverse momentum (\pte) must satisfy $\pte> 40$~\gev\ (30 \gev) for the leading (subleading) electron. 
Both electrons must be reconstructed within the range $|\eta| < 2.47$ and not in the transition region 
$1.37 < |\eta| < 1.52$ between the barrel and endcap calorimeters.  
The ATLAS \medium\ electron identification criteria~\cite{egamma_id} for 
the transverse shower shape, the longitudinal leakage into the hadronic calorimeter, and the 
association with an inner-detector track are applied to the cluster.  The electron energy
is obtained from the calorimeter measurement, and its direction is given by the associated track. A 
hit in the innermost layer of the pixel detector is required (if an active pixel layer is traversed) 
to suppress background from photon conversions. To 
suppress background from jets, the highest-\pt\ electron is required to be isolated by demanding 
that the sum of the transverse energies in the cells around the electron direction in a cone of 
radius $\Delta R=0.2$ be less than 7~\gev.  The core of the electron energy deposition is excluded and the 
sum is corrected for transverse shower leakage and pile-up from additional $pp$ collisions to make 
the isolation variable essentially independent of \pte. 
The electron trigger and reconstruction efficiencies are evaluated using tag-and-probe techniques with
$Z\rightarrow ee$ events~\cite{egamma_id} for data and MC simulation.  
Correction factors are extracted in several $\eta \times \pte$ bins and applied to the simulation.  
In cases where more than two electrons are found to satisfy the above requirements, the pair with the largest invariant mass is chosen.  
No requirement is applied to the electric charge of the electrons, as it could induce an inefficiency in the signal selection for high-$\pt$ electrons due to charge misidentification.

Each muon candidate has to be reconstructed independently in both the inner detector and the muon 
spectrometer. Its momentum is determined from a combined fit to these two 
measurements. For the $\mu^*$ search, two muon candidates with a transverse momentum (\ptmu) above 25~\gev\ are required. 
Both muons must have a minimum number of hits in the inner detector and hits in each 
of the inner, middle, and outer layers of the muon spectrometer. This requirement, 
which restricts the muon acceptance to $|\eta|<2.5$, guarantees a precise momentum measurement.
Muons with hits in the barrel--endcap 
overlap regions of the muon spectrometer ($1.05 \lesssim |\eta| \lesssim 1.4$) are discarded 
because of the limited coverage with drift-tube chambers in this angular range. 
To suppress background from cosmic rays, the muon tracks are required to 
have transverse and longitudinal impact parameters $|d_0|<0.2$~mm and $|z_0| < 1$~mm 
with respect to the selected primary vertex. To reduce background from heavy-flavour hadrons, 
each muon is required to be isolated such that $\Sigma\pt/\ptmu <5\%$, where 
the sum is over inner-detector tracks with $\pt >1$~\gev\ that are contained in a cone of radius 
$\Delta R=0.3$ surrounding the candidate muon track, the latter being excluded from the sum. 
The muon trigger and reconstruction efficiencies are evaluated using tag-and-probe techniques with
$Z\rightarrow \mu\mu$ events~\cite{Muon_reco}, and $\eta$-dependent corrections to be applied to the simulation are determined. 
The two muons are additionally required to have opposite electric charge. 
In cases where more than one pair of muons are found to satisfy the above requirements, the pair 
with the largest invariant mass is considered.

Each photon candidate is formed from a cluster of cells in the electromagnetic 
calorimeter. A photon can be reconstructed either as an unconverted photon, with no associated track, or as a photon 
that converted to an electron--positron pair, associated with one or two tracks.  
The presence of at least one photon candidate with $\ptgam > 30$~\gev\ and 
$|\eta | < 2.37$ is required in both channels. As for electrons, photons 
within the transition region between the barrel and endcap calorimeters are 
excluded. 
Photon candidates are required to satisfy the ATLAS \tight\ photon 
definition~\cite{ATLASphotons}. This selection includes constraints on the energy 
leakage into the hadronic calorimeter as well as stringent requirements on 
the energy distribution in the first and second sampling layers of the electromagnetic calorimeter. 
These requirements increase the purity of the selected photon sample by rejecting most of 
the jet background, including jets with a leading neutral hadron (usually a $\pi^0$) 
that decays into a pair of collimated photons. 
The photon-identification efficiency and shower shapes in the electromagnetic calorimeter are studied 
using FSR photons from $Z$~boson decays with loose lepton--photon separation requirements.  
Shower shapes are then adjusted in the simulation so that the 
resulting photon-identification efficiency matches the efficiency measured in data~\cite{PhotonID}.  

To further reduce background from misidentified jets, photon candidates are required to be isolated by demanding that 
either $\et^{\rm iso}<10$~\gev\ or $\et^{\rm iso}/\ptgam<1\%$, where $\et^{\rm iso}$ is
the sum of the transverse energies of the clusters within a cone of radius $\Delta R = 0.4$ surrounding the photon. 
As for the electron isolation, the clusters from the photon energy 
deposition are excluded and the sum is corrected for transverse shower leakage and pile-up. 
The relative-isolation criterion reduces the efficiency loss for high-\pt\ photons ($\ptgam>1$~\tev).
Since the photon and the leptons are expected to be well separated for the excited-lepton signal,
only photons satisfying $\Delta R(\ell,\gamma) >0.7$ are retained.  This requirement is effective at suppressing 
Drell--Yan events with FSR photons that are typically highly collimated with the leptons.  
If more than one photon candidate in an event satisfies the above requirements, the one with the 
largest \pt\ is used in the search.

Finally, two additional requirements are applied to drastically reduce the background level.
The first one, referred to as the ``$Z$ veto" in the following, requires the dilepton mass to satisfy $m_{\ell\ell} > 110$~\gev. 
The second is a variable lower bound on the dilepton-photon mass that defines the signal search region.
As a result of optimization studies, the signal region for $m_\lstar < 900$~\gev\ is 
$m_{\ell\ell\gamma} > m_\lstar + 150$~\gev.  For $m_\lstar \ge 900$~\gev, it is fixed to $m_{\ell\ell\gamma} > 1050$~\gev.
The signal efficiency for these two requirements is above $98$\% for $m_\lstar \ge 200$~\gev.

The total signal acceptance times efficiency ($A\times \epsilon$) is shown in figure~\ref{fig:sigeff} as a function of the excited-lepton mass. 
For low values of $m_\lstar$, the photon and the leptons tend to be produced more forward and have a softer \pt\ spectrum than at high mass, which explains the decrease in $A\times \epsilon$.
The lower geometrical acceptance in the muon channel is due to the requirement of hits in all three layers of precision chambers.  

\begin{figure}[ttt]
\begin{center}
\includegraphics[width=3.2truein]{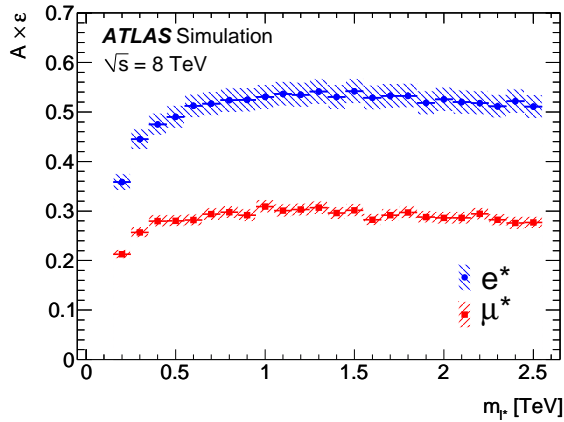}
  \caption{Acceptance times efficiency ($A\times \epsilon$) of the excited-lepton selection as a function of the excited-lepton mass ($m_\lstar$), 
evaluated for a compositeness scale of 5~\tev.  
The uncertainties correspond to the sum in quadrature of the statistical uncertainty and systematic uncertainties associated 
with the lepton and photon efficiencies.}
  \label{fig:sigeff}
\end{center}
\end{figure}


\section{Background determination}
\label{sec:bkg}


Most of the background predictions are estimated with MC samples normalized with calculated cross sections and the measured integrated luminosity of the data.  
Because the misidentification of jets as photons is not accurately modelled in the simulation, the \zpjet\ background is instead normalized to the data using 
a control region defined as $70 < m_{\ell\ell} < 110$~\gev, where the contribution from signal events is at most 3\% for $m_\lstar \ge 200$~\gev.
In this control region, the number of \zpjet\ events is estimated by subtracting from the data all simulated backgrounds except \zpjet. 
The normalization of the \zpjet\ MC sample is corrected accordingly by a scale factor, separately determined to be 
$0.53 \pm 0.10$ for both the electron and muon channels.  
The quoted uncertainty combines the statistical uncertainties on the data and simulated backgrounds and the uncertainty on the \zpgam\ cross section. 
Other sources of uncertainty including the integrated luminosity and the cross sections of the \ttbar\ and diboson processes are negligible.
Scale factors were evaluated in different \ptgam\ bins, and results are consistent within statistical uncertainties.

In the electron channel, the $W+\gamma+\rm jets$ background is also normalized to the data because of the imperfect modelling of the jet-to-electron fake rate.
For this background only, the identification criteria were relaxed for one electron to increase the MC statistics.
The $W+\gamma+\rm jets$ normalization is derived using a likelihood template fit to the data, in the same control region as for \zpjet. 
The fit is simultaneously performed on transverse mass distributions $m_{\mathrm {T}}(e_1,\met)$ and $m_{\mathrm {T}}(e_2,\met)$, 
where \met\ denotes the magnitude of the missing transverse momentum,
which is calculated~\cite{AtlasETmiss} from calorimeter cells with $|\eta|<4.9$ using the local energy calibration of electrons, photons, hadronically-decaying $\tau$-leptons and jets. 
Cells belonging to clusters not associated with such reconstructed objects as well as cells associated with a muon candidate are also included. 
The transverse mass is $m_{\mathrm {T}}(e_i,\met) = \sqrt{2 p_{\mathrm {T}}^{e_i} \met(1-\cos\Delta\phi)}$, where 
$\Delta\phi$ is the angle between the transverse momentum of electron~$i$ ($p_{\mathrm {T}}^{e_i}$) and the missing transverse momentum.
The only floating parameter in the fit is the scale factor of the $W+\gamma+\rm jets$ background,
which is found to be 
$0.22^{+0.25}_{-0.22}$~(${\rm stat} \oplus {\rm syst}$). Systematic uncertainties account for correlations between the two $m_{\mathrm {T}}$ variables, the choice of control region in which the fit is performed, the loosening 
of the electron identification criteria, and the dependence on the \zpjet\ scale factor.

The numbers of events in the control region ($70 < m_{\ell\ell} < 110$~\gev)
and the numbers after the $Z$ veto ($m_{\ell\ell}>110$~\gev) are shown in table~\ref{table-control} after scaling the \zpjet\ background, as well as the $W+\gamma+\rm jets$ background in the electron channel.  
In the control region, by construction, the total background is equal to the number of events in data.
After the $Z$ veto, the observed data are found to be consistent with the background prediction.
Good agreement is also observed between data and background in the control region 
for the lepton and photon kinematic distributions.
In particular, figure~\ref{fig:photon_pt_control} shows that the background prediction for the photon \pt\ distribution matches the data for both the $e^*$ and $\mu^*$ searches. 

\begin{table}
\caption{\label{table-control}Data yields and background expectations in the control region and after the $Z$ veto. The \zpjet\ and $W+\gamma+\rm jets$ backgrounds are scaled as described in the text.  
The uncertainties shown are purely from MC statistics, except for \zpjet\ and $W+\gamma+\rm jets$ where the statistical uncertainty on the associated scale factor is reported.}

\begin{indented}
\item[]
\begin{tabular}{@{}lcccc}
\br
                   & \multicolumn{4}{c}{Regions  (\gev)} \\ \cline{2-5}
Samples            & $70<m_{ee} < 110$  & $m_{ee} > 110$  & $70<m_{\mu\mu} < 110$ & $m_{\mu\mu} > 110$  \\
\mr
\zpgam\            & $1235\pm 25$\phantom{x}    & $208\pm 10 $   & $1067\pm 22$\phantom{x}        & $131\pm 8 $\phantom{x}    \\
\zpjet\            & $371\pm 48 $    & $25\pm 7   $    & $334\pm 43 $       & $12\pm 3$     \\  
$t\bar{t}$, diboson              & $18\pm 1$   & $19\pm 2$   & $16\pm 1$      &  \phantom{x}$6\pm 1$ \\
$W+\gamma+\rm jets$ &\phantom{x}$9\pm 9$ &\phantom{x}$21\pm 21$ & \phantom{x}- & \phantom{x}-\\
\hline
Total MC           & $1633\pm 55$\phantom{x}   & $273\pm 24$      & $1417\pm 48$\phantom{x}        & $149\pm 8$\phantom{x}    \\
\hline
Data               & 1633           &  263             & 1417               & 147           \\
\br
\end{tabular}
\end{indented}
\end{table}

\begin{figure}
\subfigure[\label{fig:photon_pt_e}]{\includegraphics[width=3.2truein]{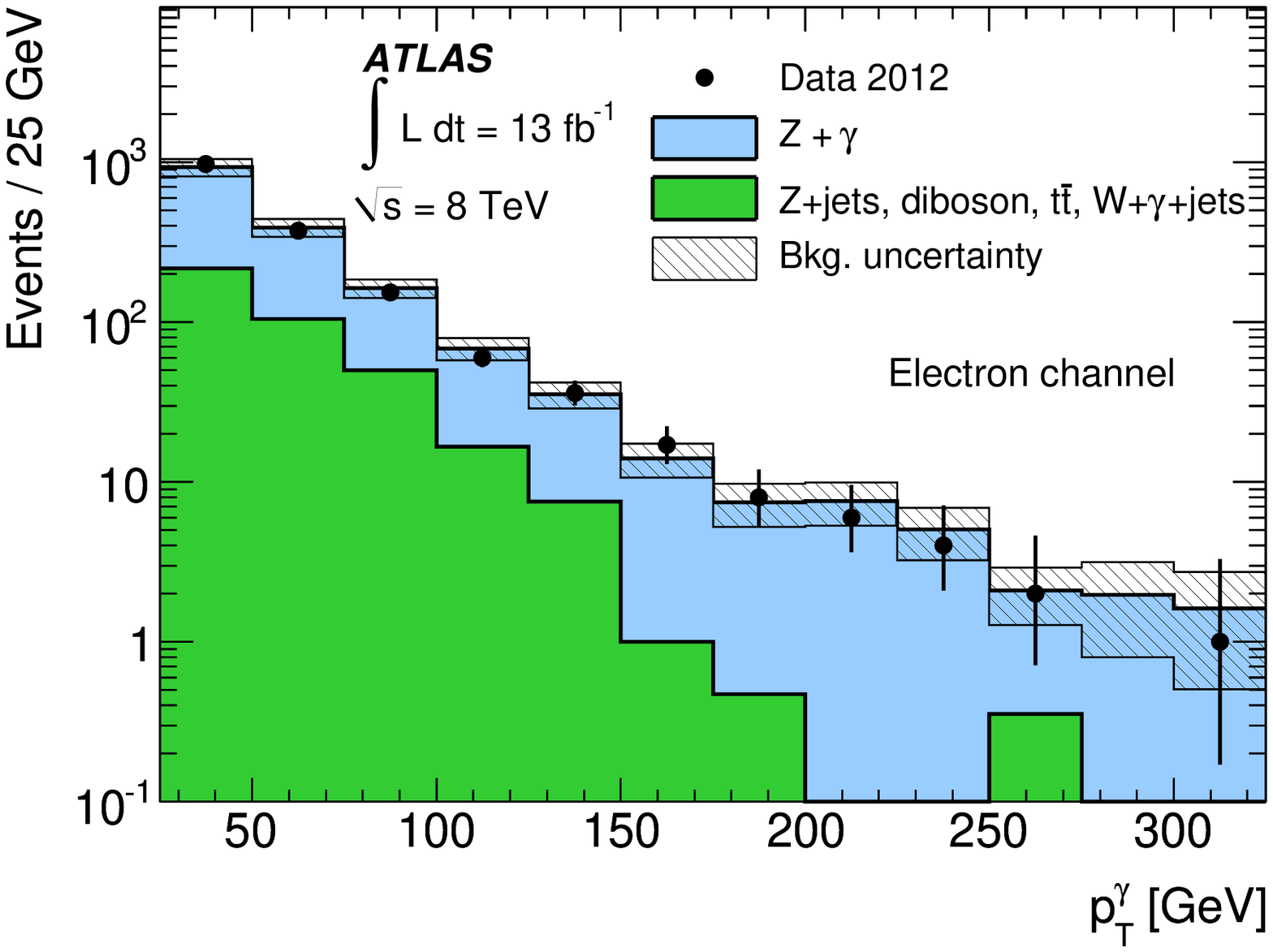}}
\subfigure[\label{fig:photon_pt_mu}]{\includegraphics[width=3.2truein]{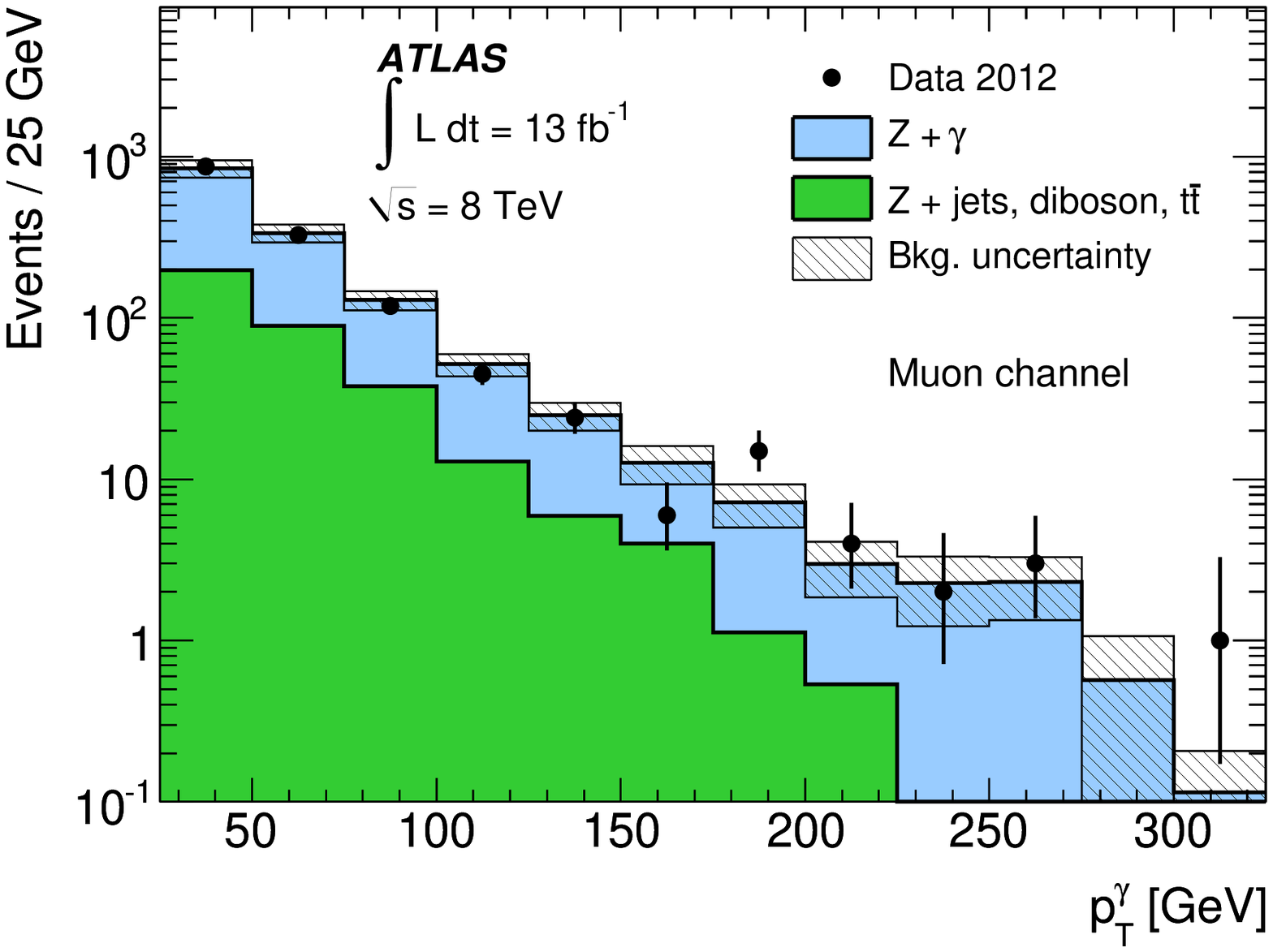}}
\caption{Distributions of the transverse momentum of the photon ($\ptgam$) for the electron~\ref{fig:photon_pt_e} and muon~\ref{fig:photon_pt_mu} 
channels, in the control region defined by the dilepton mass range $70<m_{\ell\ell} < 110$~\gev.
The background uncertainty corresponds to the sum in quadrature of the statistical uncertainties 
and the uncertainty in the data-driven \zpjet\ normalization. 
\label{fig:photon_pt_control}}
\end{figure}

Because only a small fraction of the simulated background events survive the $m_{\ell\ell} > 110$~\gev\ requirement, 
the $m_{\ell\ell\gamma}$ distributions of dominant backgrounds are separately fitted with an exponential function and extrapolated to the high-mass region.
The binned results of these fits are used as final background estimates in the statistical analysis.
The same operation is performed for the $m_{\ell\gamma}$ distribution of each background, although in this case, the fit results are not used in any numerical analysis.
The resulting background estimates are shown in figures~\ref{fig:lg-final} and \ref{fig:llg-final} as functions of the invariant mass of the $\ell\gamma$ and $\ell\ell\gamma$ systems, respectively.  
For table~\ref {table-control} and figures~\ref{fig:photon_pt_control}--\ref{fig:llg-final}, the lower bound on $m_{\ell\ell\gamma}$ described in section~\ref{sec:selection} is not applied.

\begin{figure}
\subfigure[\label{fig:eg-final}]{\includegraphics[width=3.2truein]{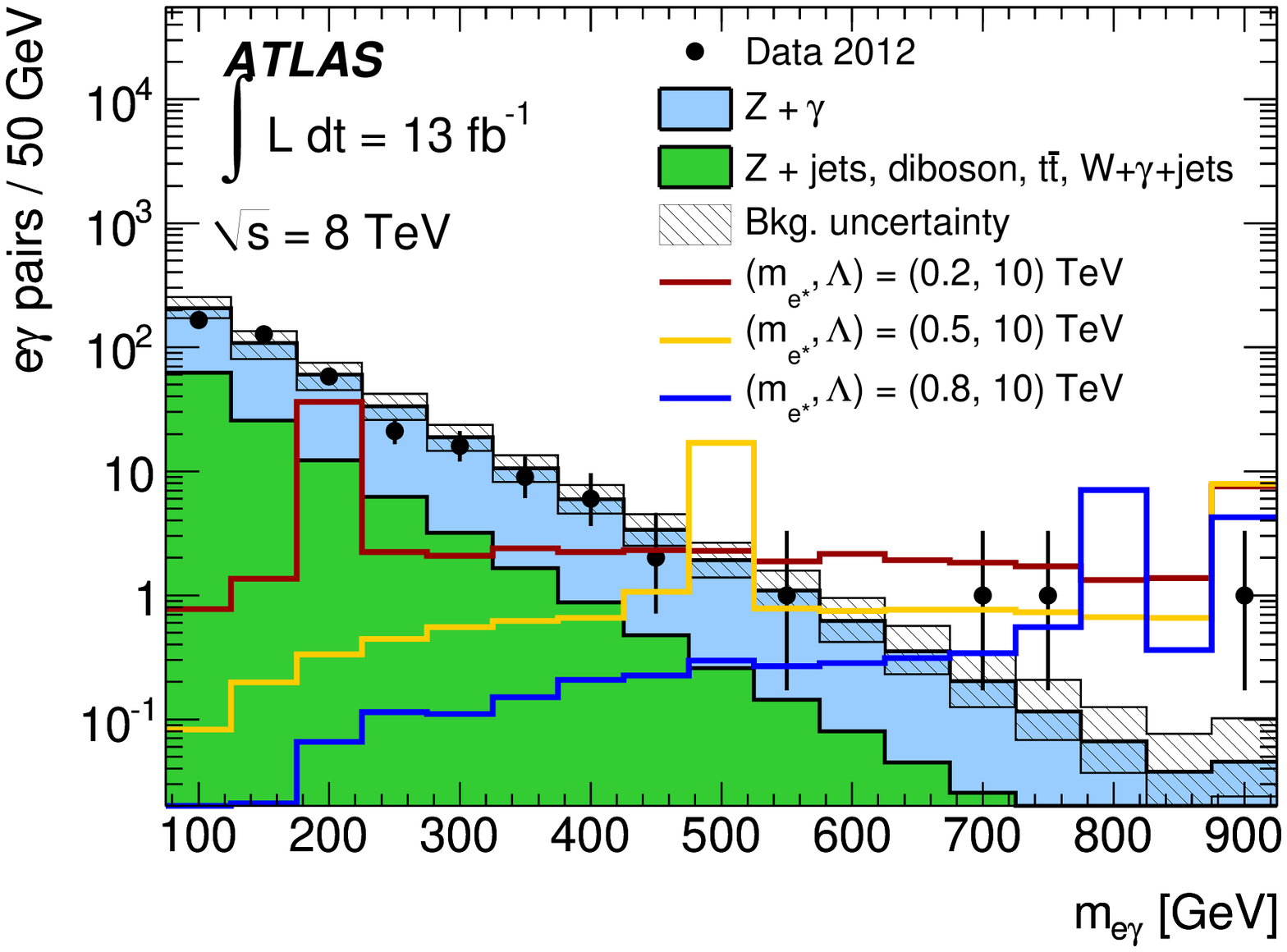}}
\subfigure[\label{fig:mug-final}]{\includegraphics[width=3.2truein]{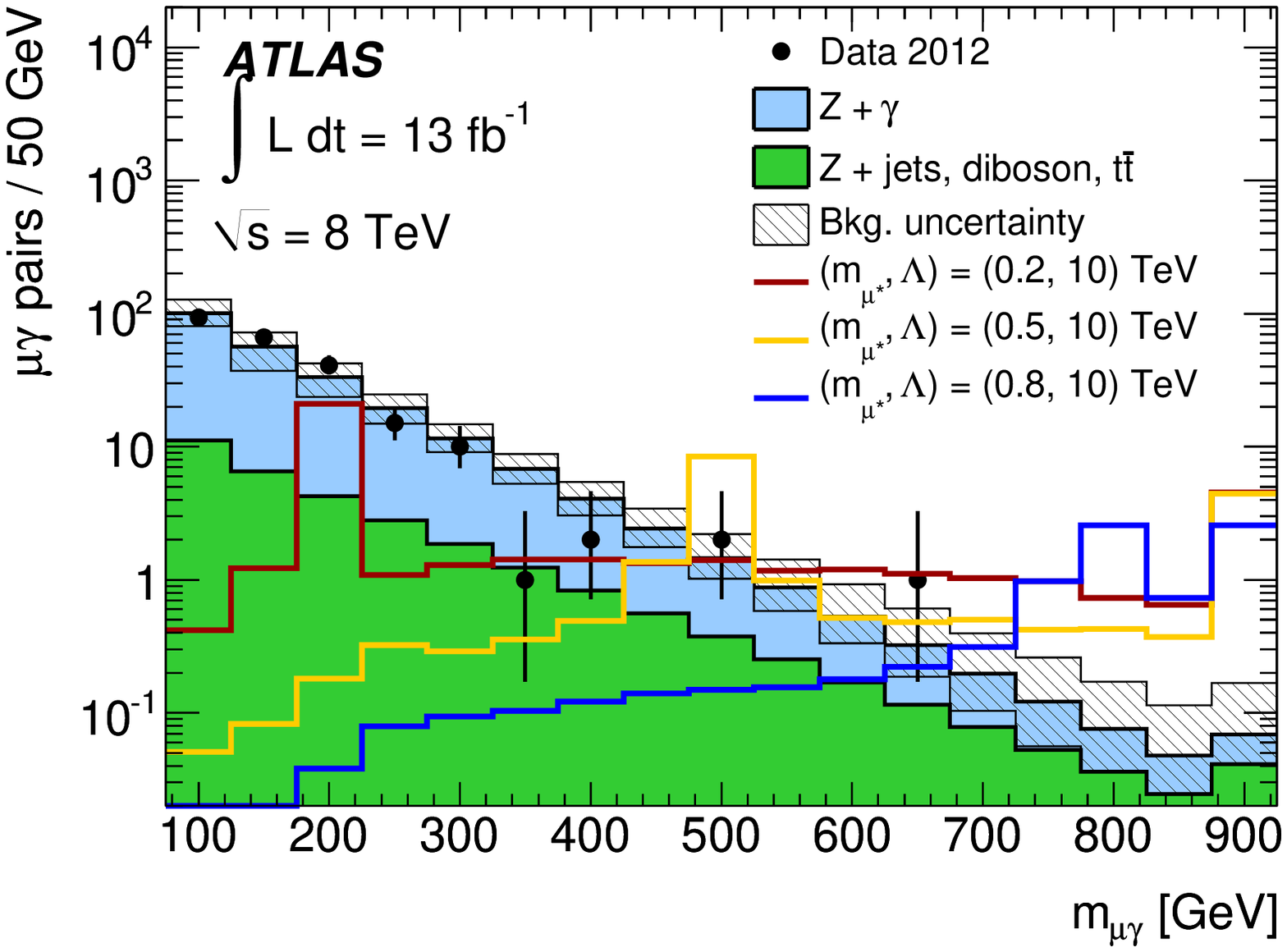}}
\caption{Distributions of the $\ell\gamma$ invariant mass ($m_{\ell\gamma}$) for the electron~\ref{fig:eg-final} and 
muon~\ref{fig:mug-final} channels after requiring the dilepton mass to satisfy $m_{\ell\ell} > 110$~\gev.
Combinations with both the leading and subleading leptons are shown.
The binned results of exponential fits are used for all backgrounds.
The background uncertainty corresponds to the sum in quadrature of the statistical and systematic uncertainties.
The last bin contains the sum of all entries with $m_{\ell\gamma}> 875$~\gev.
Signal predictions for three different values of the excited-lepton mass ($m_\lstar$) with a compositeness scale ($\Lambda$) of 10~\tev\ are also shown.
\label{fig:lg-final}}
\end{figure}

\begin{figure}
\subfigure[\label{fig:eeg-final}]{\includegraphics[width=3.2truein]{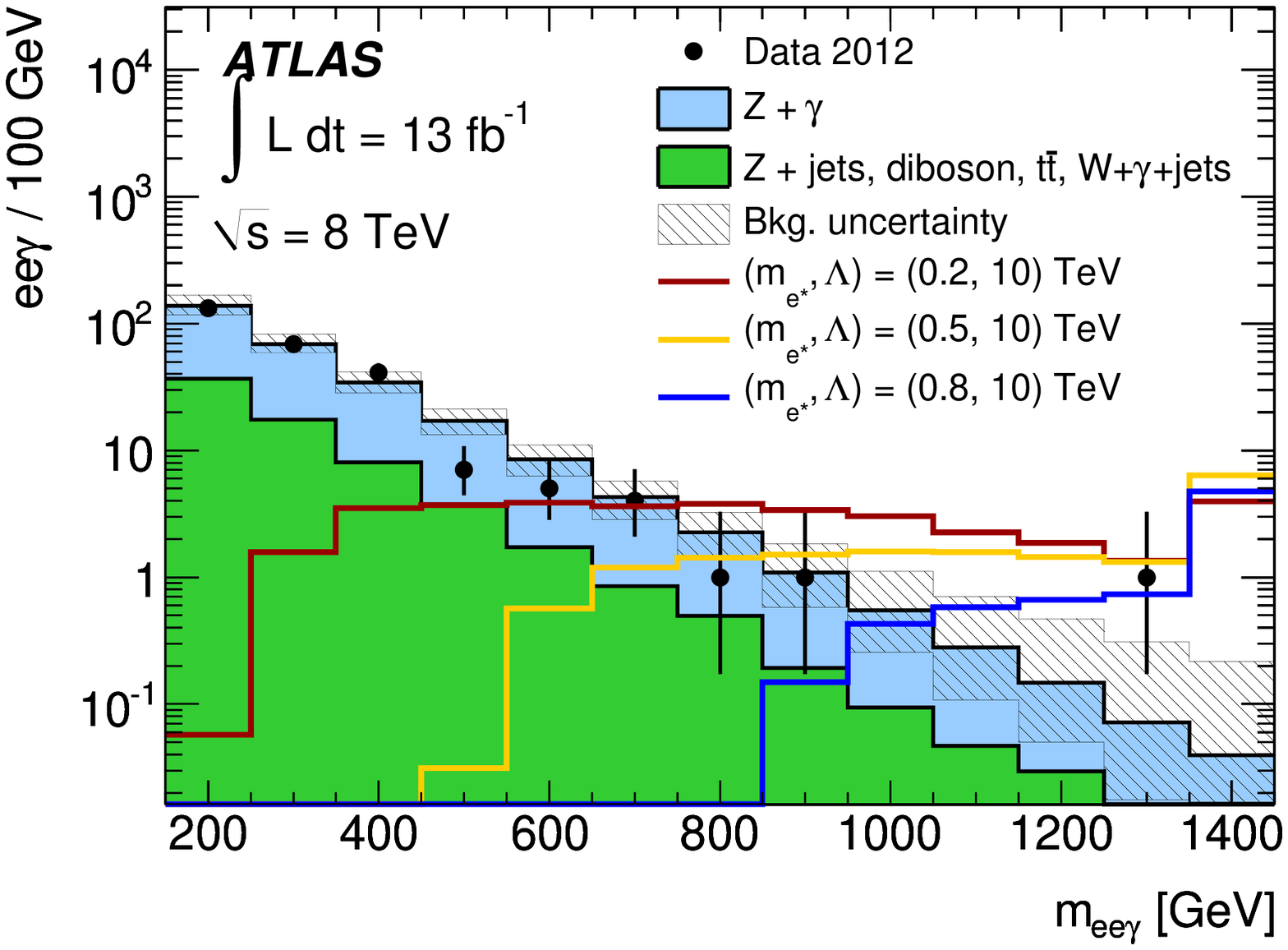}}
\subfigure[\label{fig:mumug-final}]{\includegraphics[width=3.2truein]{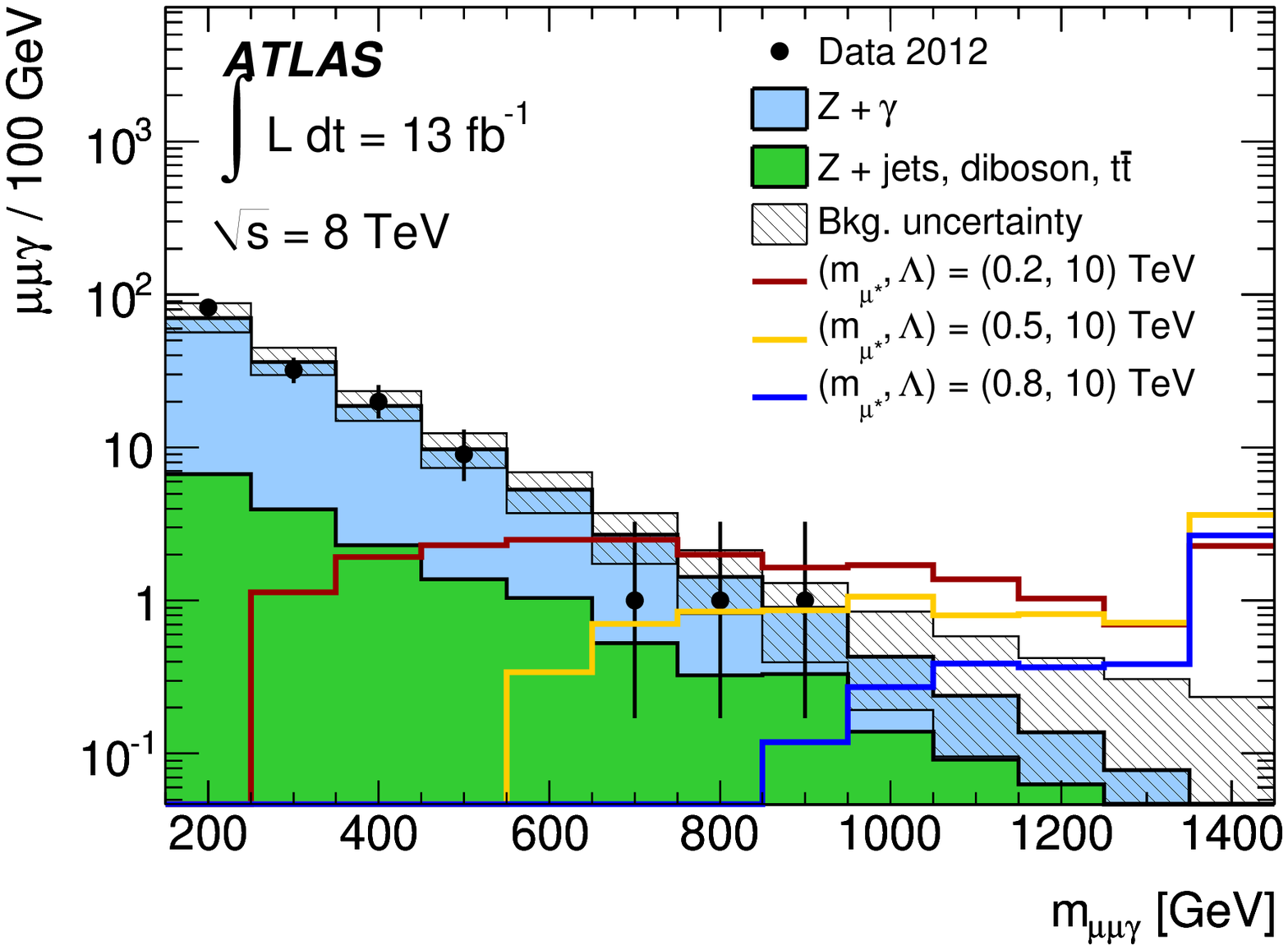}}
\caption{Distributions of the $\ell\ell\gamma$ invariant mass ($m_{\ell\ell\gamma}$) for the electron~\ref{fig:eeg-final} and muon~\ref{fig:mumug-final} 
channels after requiring the dilepton mass to satisfy $m_{\ell\ell} > 110$~\gev. 
The binned results of exponential fits are used for the \zpgam, \zpjet, \ttbar\ and $W+\gamma+\rm jets$ backgrounds.
The background uncertainty combining the statistical and systematic uncertainties is displayed as the hatched area. 
The last bin contains the sum of all events with $m_{\ell\ell\gamma}> 1350$~\gev.
Signal predictions for three different values of the excited-lepton mass ($m_\lstar$) with a compositeness scale ($\Lambda$) of 10~\tev\ are also shown.
\label{fig:llg-final}}
\end{figure}


\section{Systematic uncertainties}


The most important sources of uncertainty are discussed below and summarized in table~\ref{tab:systematics1000}. 
A large part of the background uncertainty comes from the \zpgam\ cross-section calculation.  
It includes the renormalization and factorization scale uncertainties, obtained by varying independently each scale by a factor of two, 
as well as uncertainties in the PDFs and the strong coupling constant $\alpha_s$. These 
uncertainties are evaluated by generating \zpgam\ \sherpa\ samples for the 52 CT10 eigenvector PDF sets, 
the four CT10.AS PDF sets corresponding to $\alpha_s=0.116,0.117,0.119,0.120$, and the four combinations of scales.  
For $m_{\ell\ell\gamma}>350$~\gev\ ($m_{\ell\ell\gamma} > 1050$~\gev), the resulting uncertainty 
is $^{+25\%}_{-16\%}$ $\left(^{+32\%}_{-18\%}\right)$ for both channels. 
Cross-section uncertainties for the $\ttbar$ and diboson processes have a negligible impact on the total background uncertainty. 

The statistical uncertainties associated with the $m_{\ell\ell\gamma}$ fits contribute to the background uncertainty at a comparable level at low mass, and become increasingly important at high mass. 
The sum in quadrature of fit uncertainties, including uncertainties on data-driven scale factors for the relevant backgrounds, increases from about $\pm 20$\% for $m_{\ell\ell\gamma}>350$~\gev\ in 
both channels to $^{+215\%}_{-65\%}$ $\left(^{+200\%}_{-60\%}\right)$ for $m_{\ell\ell\gamma} > 1050$~\gev\ in the $e^*$ ($\mu^*$) search.
The main contributions come from the \zpgam\ and \zpjet\ backgrounds, as well as the $W+\gamma+\rm jets$ background in the electron channel.

Experimental systematic uncertainties that affect both the signal and background yields include the 
uncertainty on the luminosity measurement and uncertainties in particle 
reconstruction and identification as described below.

The uncertainty on the integrated luminosity is 2.8\%. It is based 
on a preliminary calibration of the luminosity scale derived from beam-separation scans~\cite{lumi7TeV} performed in November 2012.

The total uncertainty on the photon reconstruction and identification efficiencies is 4\%~\cite{PhotonID}.
The combination of uncertainties on the electron trigger, reconstruction, identification and isolation efficiencies results in a 2\% uncertainty on both the 
signal efficiency and background level.  
The combined uncertainty on the trigger, reconstruction and identification efficiencies for muons is
estimated to increase linearly as a function of $m_\lstar$ to about 2\% for $m_\lstar =2$~\tev. 
This uncertainty is dominated by the impact of large energy loss from muon bremsstrahlung in the calorimeter.  
The sum in quadrature of the lepton and photon uncertainties for the lowest $m_{\ell\ell\gamma}$ threshold is shown in table~\ref{tab:systematics1000}.
Uncertainties on the energy scale and resolution for final-state objects have a negligible effect on signal and background selection efficiencies.

The impact of the \lstar\ decay width on the signal selection efficiency was also investigated.  
The decay width is computed with formulas given in~\cite{BaurSpiraZerwas}. It increases with $m_\lstar$ and decreases with $\Lambda$, 
and over the $\Lambda$--$m_\lstar$ region accessible in these searches, it ranges from $\simeq 1 \mev$ to $\simeq200\gev$. 
Signal efficiencies were computed at the generator level for different values of $\Lambda$, and
efficiency variations were observed to be at most 1\%, which is 
negligible compared to the other uncertainties in the selection efficiency.  

\begin{table}[!htbp]
\caption{\label{tab:systematics1000}Dominant uncertainties on the expected numbers of events for the lowest-mass search region, $m_{\ell\ell\gamma}>350$~\gev.
The theory uncertainty reported for the background corresponds to the uncertainty on the \zpgam\ cross section only.}
\begin{indented}
\item[]
\setlength{\extrarowheight}{1pt}
\begin{tabular}{@{}lcccc}
\br
Source          & \multicolumn{2}{c}{$e^*$}  & \multicolumn{2}{c}{$\mu^*$} \\ \cline{2-5}
                & Signal       & Background    & Signal       & Background  \\
\mr
Theory          & 1\%           & $^{+25\%}_{-16\%}$\phantom{xx}            & 1\%           & $^{+25\%}_{-16\%}$\phantom{xx}    \\
Statistics   & -           & 18\%\phantom{x}          & -           & 21\%\phantom{x}  \\
Luminosity      & 3\%          & 3\%           & 3\%          & 3\%    \\
Efficiencies      & 5\%          & 5\%           & 5\%          & 5\%    \\
\br
\end{tabular}
\end{indented}
\end{table}


\section{Results}
\label{sec:results}


The $m_{\ell\ell\gamma}$ distributions are shown in figure~\ref{fig:llg-final} for the data, the expected backgrounds, and three signal predictions.
The expected and observed numbers of events in each of the search regions, used for the statistical analysis, are shown in tables~\ref{tab:yields_e} and~\ref{tab:yields_mu} for 
the electron and muon channels, respectively. 
The uncertainties include both the statistical and systematic contributions as described earlier.
The data are consistent with the background expectation, and no significant excess is observed in the signal region.

An upper limit on the cross section times branching 
ratio $\sigma(pp\to\ell\ell^*)\times B(\lstar \rightarrow \ell\gamma)$ is determined for each channel and each $m_{\lstar}$ hypothesis
at the 95\% credibility level (CL) using a Bayesian approach~\cite{bayesianMethod} with a flat positive 
prior for $\sigma B$. Systematic uncertainties are incorporated into the limit calculation as nuisance parameters with 
Gaussian priors.  Uncertainties in particle reconstruction and identification efficiencies as well as the uncertainty on 
the luminosity are fully correlated between signal and backgrounds.  All other uncertainties are uncorrelated. 
The expected limit is evaluated as the median of the upper-limit distribution obtained with a set of background-only pseudo-experiments.
Figure~\ref{fig:limits1d} shows the 95\% CL expected and observed limits on 
$\sigma B$ for the $e^{*}$ and $\mu^{*}$ searches.  For $m_{\lstar}\ge 800$~\gev, 
the observed upper limits are 0.75~fb and 0.90~fb for the electron and muon channels, respectively.  
The sensitivity to the prior for $\sigma B$ was studied using a reference prior\cite{Diego}, resulting in 20--25\% better limits for both channels. 
Theoretical predictions of $\sigma B$ for three different values of $\Lambda$ 
are also displayed in figure~\ref{fig:limits1d}, along with the uncertainties from 
renormalization and factorization scales and PDFs. These uncertainties are 
shown for illustrative purpose only and are not used when setting limits.  

For each $m_{\lstar}$ hypothesis, the limit on $\sigma B$ is then translated into a lower bound on the compositeness scale.
This bound corresponds to the value of $\Lambda$ for which the theoretical prediction $\sigma B(m_{\lstar},\Lambda)$ is equal to the upper limit on $\sigma B$.
The excluded region in the $\Lambda$--$m_{\lstar}$ plane is shown in figure~\ref{fig:limits2d} for both the $e^{*}$ and $\mu^{*}$ searches.
For $m_{\lstar}=\Lambda$, excited-electron and excited-muon masses are both excluded at 95\% CL up to 2.2~\tev.
The limits obtained with $\sqrt{s}=7$~\tev\ data by ATLAS~\cite{lstar2011} and CMS~\cite{lstarCMS} are also shown.
 
\begin{table}[!htbp]
\caption{Data yields and background expectation as a function of a lower bound on $m_{ee\gamma}$ for the $e^*$ search.
The uncertainties represent the sum in quadrature of the statistical and systematic uncertainties.
\label{tab:yields_e}}
\begin{indented}
\item[]
\setlength{\extrarowheight}{1pt}
\begin{tabular}{@{}cccc}
\br
$m_{ee\gamma}$ region (\gev) & \zpgam\ & Total bkg & Data \\
\mr
$> 350$     & $53^{+16}_{-14}$    &  $\;69 ^{+18}_{-16}\;$   & 60     \\
$> 450$     &  $27 \pm 9\;$    &  $\;34 ^{+11}_{-10}\;$     & 19    \\
$> 550$     &  $14 ^{+6}_{-5}\;$      &   $17 ^{+7}_{-6}\;$     & 12      \\
$> 650$     &  $7.0 ^{+3.6}_{-3.3}$   &  $8.7 ^{+4.8}_{-3.4}$  & \phantom{x}7     \\
$> 750$     &  $3.5 ^{+2.5}_{-1.9}$  &  $4.4 ^{+3.4}_{-2.0}$  & \phantom{x}3       \\
$> 850$     &  $1.8 ^{+1.8}_{-1.1}$   &  $2.2 ^{+2.5}_{-1.1}$  & \phantom{x}2      \\
$> 950$     &   $0.9 ^{+1.2}_{-0.6}$   &  $1.1 ^{+1.7}_{-0.6}$  & \phantom{x}1       \\
$> 1050$\phantom{x}    &  $0.4 ^{+0.8}_{-0.3}$   &  $0.5^{+1.2}_{-0.4}$  & \phantom{x}1     \\
\br
\end{tabular}
\end{indented}
\end{table}

\begin{table}[!htbp]
\caption{Data yields and background expectation as a function of a lower bound on $m_{\mu\mu\gamma}$ for the $\mu^*$ search.
The uncertainties represent the sum in quadrature of the statistical and systematic uncertainties.
\label{tab:yields_mu}}
\begin{indented}
\item[]
\setlength{\extrarowheight}{1pt}
\begin{tabular}{@{}cccc}
\br
$m_{\mu\mu\gamma}$ region (\gev) & \zpgam\ & Total bkg & Data \\
\mr
$> 350$     &  $33^{+11}_{-9}$     &  $40 ^{+11}_{-10}$     & 32   \\
$> 450$     &  $\;17 \pm 6$\phantom{x}    &  $21 ^{+7}_{-6}\;$     & 12   \\
$> 550$     &  $8.7 ^{+3.8}_{-3.6}$   &  $11 ^{+5}_{-4}\;$     & \phantom{x}3    \\
$> 650$     &  $4.4 ^{+2.4}_{-2.2}$ &  $5.9 ^{+3.5}_{-2.6}$   & \phantom{x}3    \\
$> 750$     & $2.2 ^{+1.5}_{-1.3}$   &  $3.2 ^{+2.6}_{-1.6}$   &\phantom{x}2    \\
$> 850$     &  $1.1 ^{+1.0}_{-0.8}$   &  $1.7 ^{+2.0}_{-0.9}$   & \phantom{x}1    \\
$> 950$     &  $0.6 ^{+0.7}_{-0.4}$   &  $0.9 ^{+1.5}_{-0.6}$   & \phantom{x}0    \\
$> 1050$\phantom{x}      & $0.3 ^{+0.5}_{-0.2}$   &  $0.5 ^{+1.0}_{-0.3}$   &\phantom{x}0    \\
\br
\end{tabular}
\end{indented}
\end{table}

\begin{figure}[ht!]
\subfigure[\label{fig:limits1d-e}]{\includegraphics[width=3.2truein]{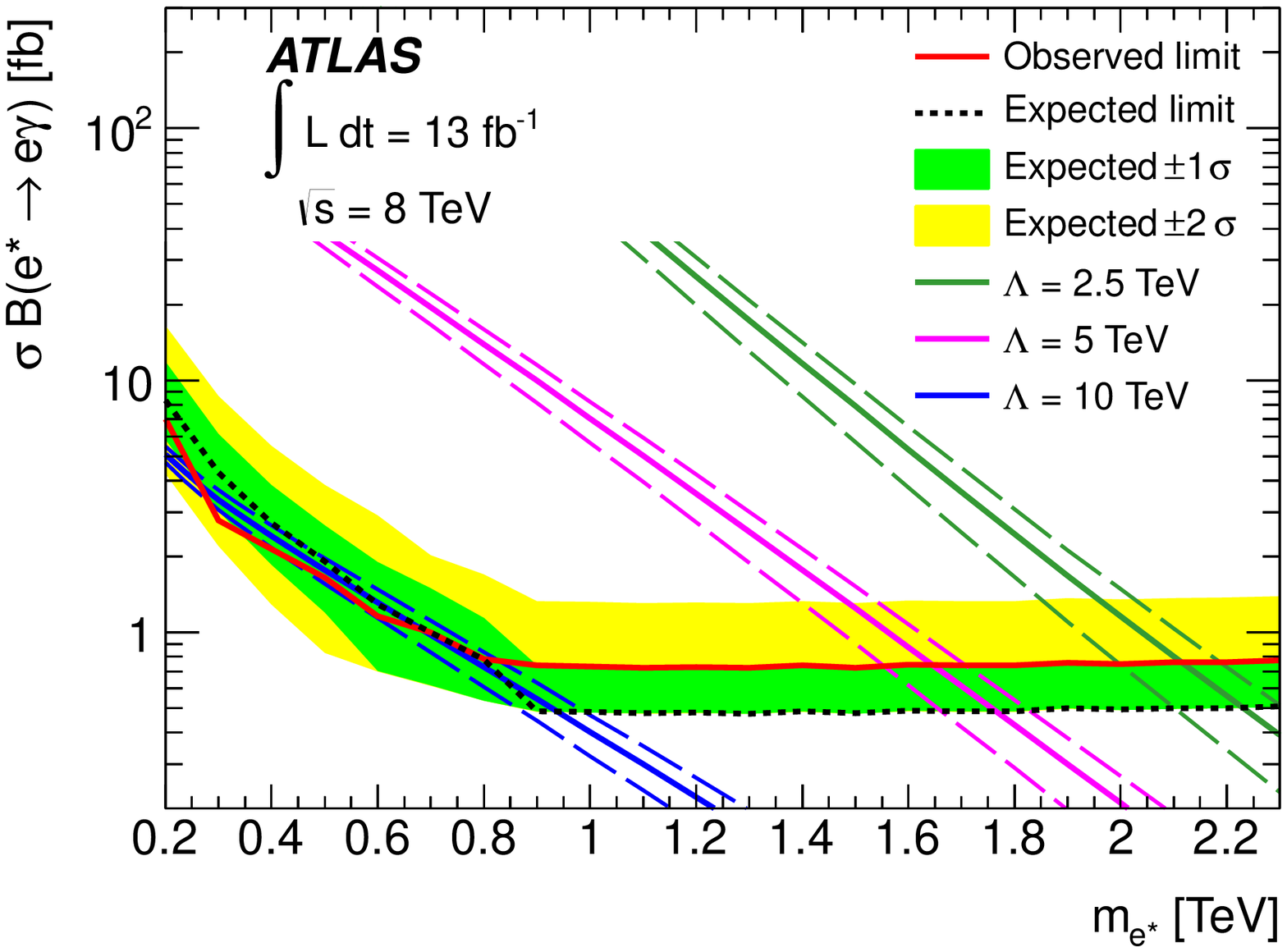}}
\subfigure[\label{fig:limits1d-mu}]{\includegraphics[width=3.2truein]{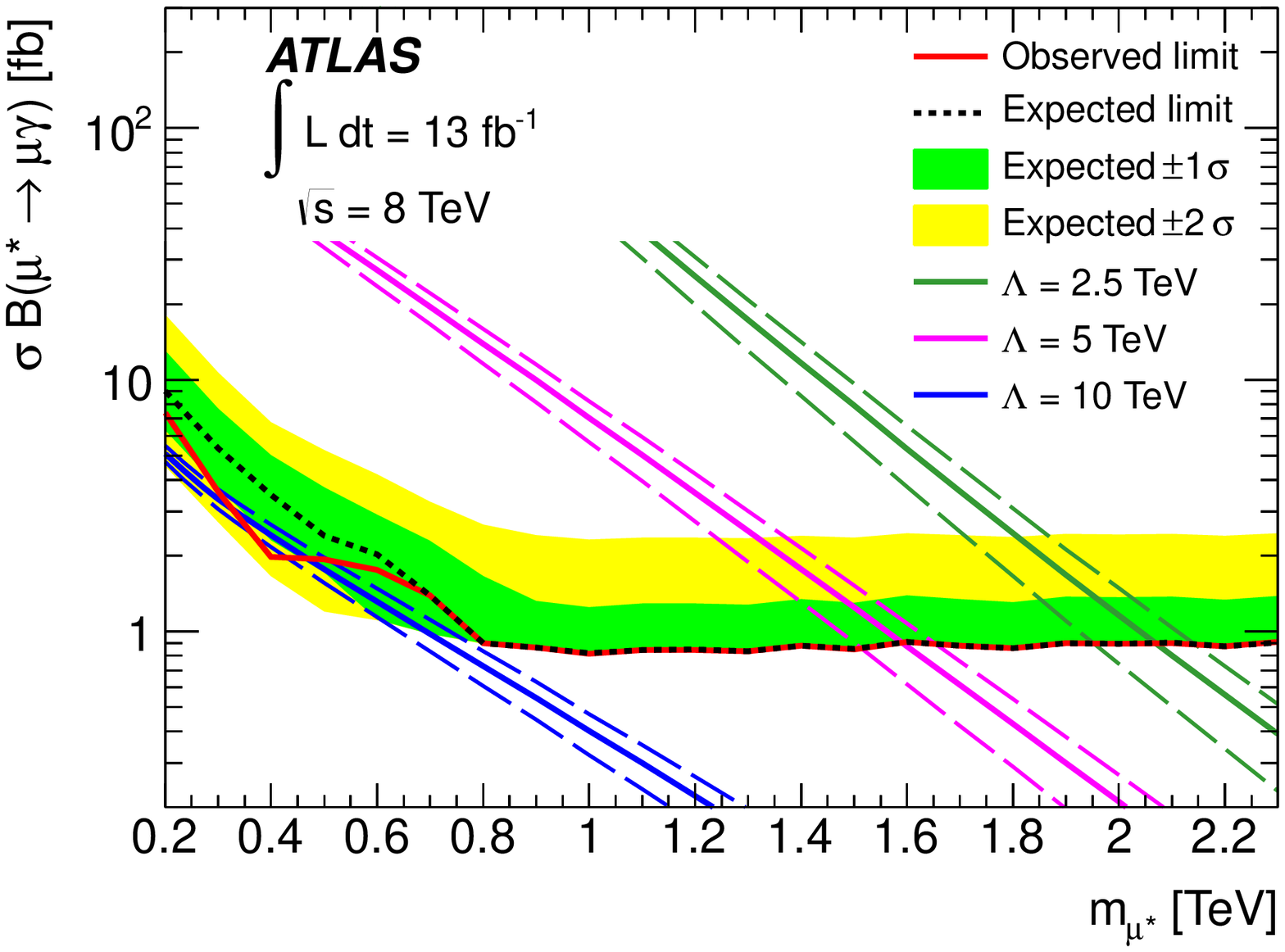}}
\caption{Upper limits at 95\% CL on the cross section times branching ratio ($\sigma B$) as a function of the excited-lepton mass ($m_{\lstar}$), for the electron~\ref{fig:limits1d-e}
and muon~\ref{fig:limits1d-mu} channels. LO signal predictions with uncertainties from renormalization and factorization scales and PDFs 
are shown for three different compositeness scales ($\Lambda$).
\label{fig:limits1d}}
\end{figure}

\begin{figure}[hbpt]
\subfigure[\label{fig:limits2d-e}]{\includegraphics[width=3.2truein]{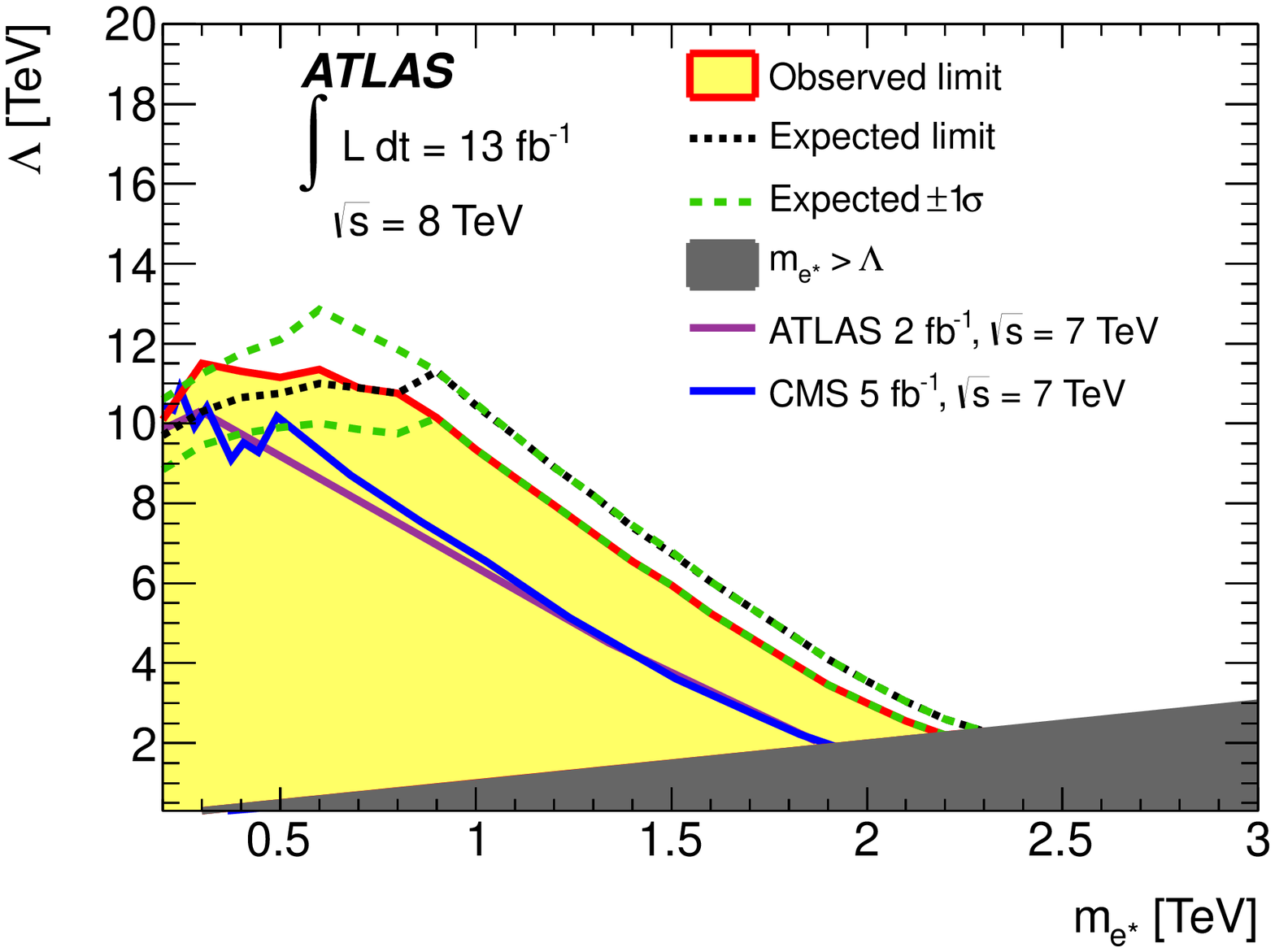}}
\subfigure[\label{fig:limits2d-mu}]{\includegraphics[width=3.2truein]{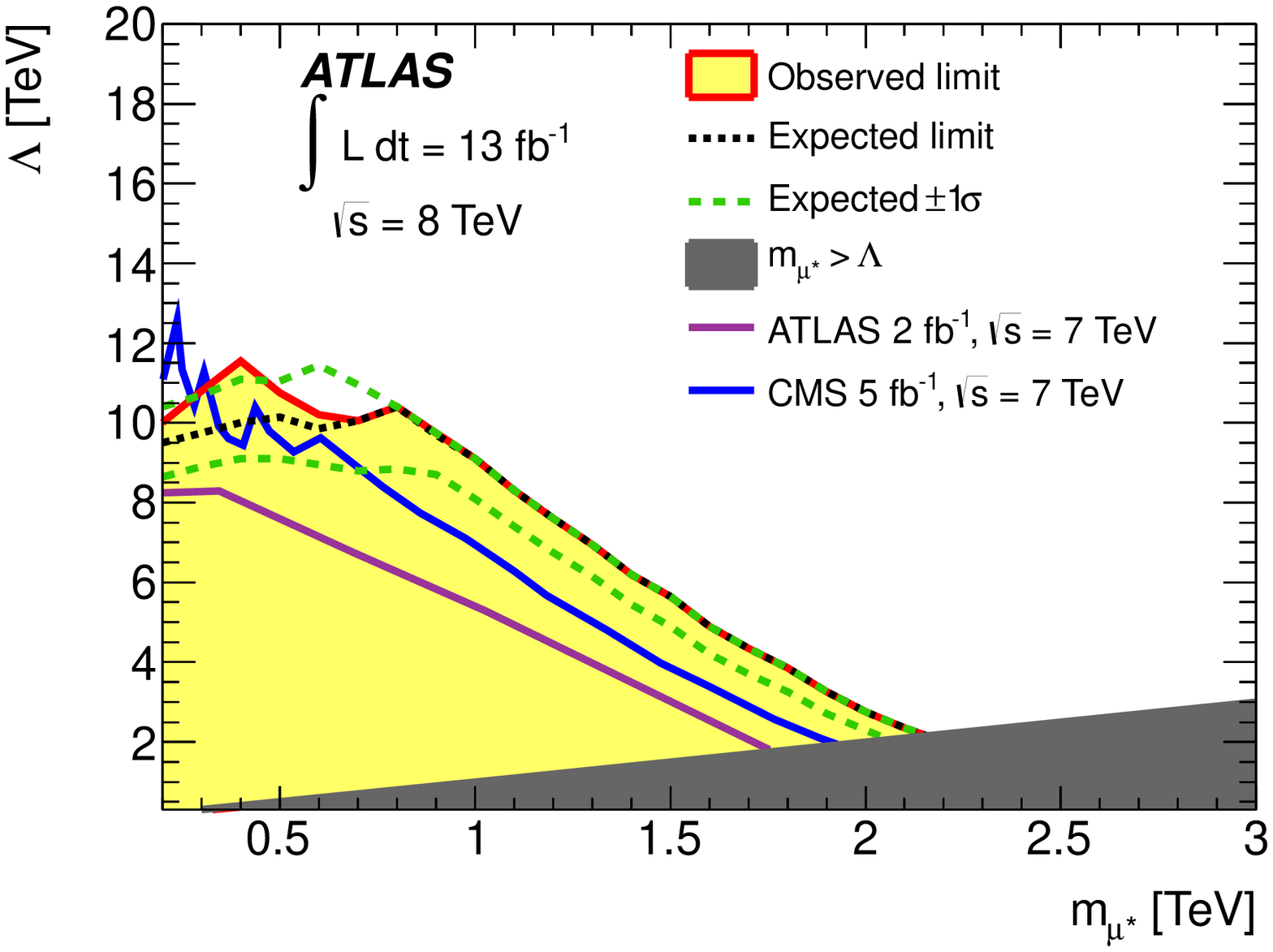}}
\caption{Exclusion limits in the compositeness scale ($\Lambda$) vs excited-lepton mass~($m_{\lstar}$) parameter space for the electron~\ref{fig:limits2d-e} and muon~\ref{fig:limits2d-mu} channels.
The filled area is excluded at 95\% CL. No limits are set in the dark shaded region $m_{\lstar} > \Lambda$ where the model is not applicable.
\label{fig:limits2d}}
\end{figure}


\section{Conclusions}


The results of a search for excited electrons and excited muons with the ATLAS detector at the LHC are reported, 
using a sample of $\sqrt{s}=8$~\tev\ $pp$ collisions corresponding to an integrated 
luminosity of 13~\ifb.  The observed data are consistent with SM background expectations.  
An upper limit is set at 95\% CL on the cross section times branching ratio $\sigma 
B(\lstar \rightarrow \ell\gamma)$ as a function of the excited-lepton mass.  
For $m_{\ell^*}\geq 0.8$~\tev, the respective limits on $\sigma B$ are 0.75~fb and 0.90~fb for the $e^*$ and $\mu^*$ searches.
These upper limits are converted into lower bounds on the compositeness scale $\Lambda$.
In the special case where $\Lambda = m_{\lstar}$, excited-electron and excited-muon masses below 2.2~\tev\ are  excluded. 


\input{Acknowledgements}



\clearpage
\onecolumn
\input{atlas_authlist}
\clearpage


\end{document}

%% file: Acknowledgements.tex



\section*{Acknowledgements}

We thank CERN for the very successful operation of the LHC, as well as the
support staff from our institutions without whom ATLAS could not be
operated efficiently.

We acknowledge the support of ANPCyT, Argentina; YerPhI, Armenia; ARC,
Australia; BMWF and FWF, Austria; ANAS, Azerbaijan; SSTC, Belarus; CNPq and FAPESP,
Brazil; NSERC, NRC and CFI, Canada; CERN; CONICYT, Chile; CAS, MOST and NSFC,
China; COLCIENCIAS, Colombia; MSMT CR, MPO CR and VSC CR, Czech Republic;
DNRF, DNSRC and Lundbeck Foundation, Denmark; EPLANET, ERC and NSRF, European Union;
IN2P3-CNRS, CEA-DSM/IRFU, France; GNSF, Georgia; BMBF, DFG, HGF, MPG and AvH
Foundation, Germany; GSRT and NSRF, Greece; ISF, MINERVA, GIF, DIP and Benoziyo Center,
Israel; INFN, Italy; MEXT and JSPS, Japan; CNRST, Morocco; FOM and NWO,
Netherlands; BRF and RCN, Norway; MNiSW, Poland; GRICES and FCT, Portugal; MERYS
(MECTS), Romania; MES of Russia and ROSATOM, Russian Federation; JINR; MSTD,
Serbia; MSSR, Slovakia; ARRS and MIZ\v{S}, Slovenia; DST/NRF, South Africa;
MICINN, Spain; SRC and Wallenberg Foundation, Sweden; SER, SNSF and Cantons of
Bern and Geneva, Switzerland; NSC, Taiwan; TAEK, Turkey; STFC, the Royal
Society and Leverhulme Trust, United Kingdom; DOE and NSF, United States of
America.

The crucial computing support from all WLCG partners is acknowledged
gratefully, in particular from CERN and the ATLAS Tier-1 facilities at
TRIUMF (Canada), NDGF (Denmark, Norway, Sweden), CC-IN2P3 (France),
KIT/GridKA (Germany), INFN-CNAF (Italy), NL-T1 (Netherlands), PIC (Spain),
ASGC (Taiwan), RAL (UK) and BNL (USA) and in the Tier-2 facilities
worldwide.

%% file: atlas_authlist.tex
\begin{flushleft}
{\Large The ATLAS Collaboration}

\bigskip

G.~Aad$^{\rm 48}$,
T.~Abajyan$^{\rm 21}$,
B.~Abbott$^{\rm 112}$,
J.~Abdallah$^{\rm 12}$,
S.~Abdel~Khalek$^{\rm 116}$,
O.~Abdinov$^{\rm 11}$,
R.~Aben$^{\rm 106}$,
B.~Abi$^{\rm 113}$,
M.~Abolins$^{\rm 89}$,
O.S.~AbouZeid$^{\rm 159}$
% ...
\bigskip
\\
% ...
$^{11}$ Institute of Physics, Azerbaijan Academy of Sciences, Baku, Azerbaijan\\
$^{12}$ Institut de F{\'\i}sica d'Altes Energies and Departament de F{\'\i}sica de la Universitat Aut{\`o}noma de Barcelona, Barcelona, Spain\\
% ...
$^{21}$ Physikalisches Institut, University of Bonn, Bonn, Germany\\
% ...
$^{48}$ Fakult{\"a}t f{\"u}r Mathematik und Physik, Albert-Ludwigs-Universit{\"a}t, Freiburg, Germany\\
% ...
$^{89}$ Department of Physics and Astronomy, Michigan State University, East Lansing MI, United States of America\\
% ...
$^{106}$ Nikhef National Institute for Subatomic Physics and University of Amsterdam, Amsterdam, Netherlands\\
% ...
$^{112}$ Homer L. Dodge Department of Physics and Astronomy, University of Oklahoma, Norman OK, United States of America\\
$^{113}$ Department of Physics, Oklahoma State University, Stillwater OK, United States of America\\
% ...
$^{116}$ LAL, Universit{\'e} Paris-Sud and CNRS/IN2P3, Orsay, France\\
% ...
$^{159}$ Department of Physics, University of Toronto, Toronto ON, Canada\\
% ...
\end{flushleft}
